\documentclass[]{pasj02}
\usepackage{ulem}
\usepackage{tabularx}
\usepackage{threeparttable}
\usepackage{natbib}
\usepackage{siunitx}
\usepackage[switch,mathlines]{lineno}
\Received{$\langle$reception date$\rangle$}
\Accepted{$\langle$acception date$\rangle$}
\Published{$\langle$publication date$\rangle$}
\begin{document}

\title{X-ray Study on Propagation of Non-thermal Particles in Microquasar SS~433/W~50 Extended Jets}
\author{Kazuho \textsc{Kayama},\altaffilmark{1, 2}$^{*}$ 
Takaaki  \textsc{Tanaka},\altaffilmark{3, 1}
Hiroyuki \textsc{Uchida},\altaffilmark{1}
Takeshi Go \textsc{Tsuru},\altaffilmark{1} 
Yoshiyuki \textsc{Inoue},\altaffilmark{4, 5, 6}
Dmitry \textsc{Khangulyan},\altaffilmark{7, 8}
Naomi \textsc{Tsuji},\altaffilmark{10, 5, 9}
Hiroaki \textsc{Yamamoto}\altaffilmark{11}
}%
\altaffiltext{1}{Department of Physics, Kyoto University, Kitashirakawa Oiwake-cho, Sakyo, Kyoto 606-8502, Japan}
\altaffiltext{2}{Remote Sensing Technology Center of Japan, 3-17-1, Toranomon, Minato-ku, Tokyo, 105-0001, Japan}
\altaffiltext{3}{Department of Physics, Konan University, 8-9-1 Okamoto, Higashinada, Kobe, Hyogo 658-8501, Japan}
\altaffiltext{4}{Department of Earth and Space Science, Graduate School of Science, Osaka University, Toyonaka, Osaka 560-0043, Japan}
\altaffiltext{5}{Interdisciplinary Theoretical \& Mathematical Science Program (iTHEMS), RIKEN, 2-1 Hirosawa, Saitama 351-0198, Japan}
\altaffiltext{6}{Kavli Institute for the Physics and Mathematics of the Universe (WPI), UTIAS, The University of Tokyo, Kashiwa, Chiba 277-8583, Japan}
\altaffiltext{7}{Key Laboratory of Particle Astrophysics, Institute of High Energy Physics, Chinese Academy of  Sciences, 100049 Beijing, People's Republic of China}
\altaffiltext{8}{Tianfu Cosmic Ray Research Center, 610000 Chengdu, Sichuan, People's Republic of China}
\altaffiltext{9}{Department of Physics, Rikkyo University, Nishi-Ikebukuro 3-34-1, Toshima-ku, Tokyo 171-8501, Japan}
\altaffiltext{10}{Faculty of Science, Kanagawa University, 3-27-1 Rokukakubashi, Kanagawa-ku, Yokohama-shi, Kanagawa 221-8686, Japan}
\altaffiltext{11}{Department of Astrophysics, Nagoya University, Chikusa-ku, Nagoya, Aichi 464-8602, Japan}

\email{kayama.kazuho.57r@kyoto-u.jp}


\maketitle

\begin{abstract}
SS~433, located at the center of the W~50 radio nebula, is a binary system that ejects jets oriented east-west with precessional motion.
X-ray lobes, containing compact "knots" labeled as head (e1), lenticular (e2), and ring (e3) in the east, as well as w1, w1.5, and w2 in the west, have been detected along the jets directions.
Very-high-energy $\gamma$-ray emission has also been detected from regions containing these X-ray knots, suggesting highly efficient particle acceleration in the jets.
In our previous study, we performed X-ray imaging spectroscopy of the western lobe of W~50 to investigate spectral variations.
%We found that the X-ray spectrum gradually steepens from w1 to w2, and rapidly softens immediately outside of w2.
In this work, we extend our study to the eastern region using XMM-Newton observations to provide a more comprehensive picture of the X-ray emission from the SS~433 jets.
Our results show no detectable synchrotron emission between SS~433 and the innermost knot (head).
We also found that the X-ray spectrum of the eastern jet gradually steepens as one moves away from SS~433. 
While a similar spectral evolution is observed in the western jet, there are also noticeable differences. 
In the western lobe, the spectrum initially gradually steepens and then undergoes an abrupt softening outside the knot w2. 
However, in the eastern jet, no such rapid steepening is observed at the lenticular knot, which corresponds to w2 in the west. 
Furthermore, the observed brightening and spectral variations in the eastern jet cannot be explained by simply adjusting the parameters of the model used for the western side, suggesting the involvement of additional physical processes such as particle re-acceleration.
\end{abstract}
%\pagewiselinenumbers

\section{Introduction}
SS~433 is a Galactic microquasar consisting of a black hole or neutron star and an optical companion star \citep{Hillwig_2014}.
It ejects east--west jets that exhibit precessional motion with a period of $\sim$ 163 days \citep{Margon_1989}.
SS~433 is located almost at the center of the giant Galactic radio nebula W~50, which consists of a shell with a radius of \timeform{29'} and asymmetric ``wings''.

Two main scenarios have been proposed to explain the unique morphology of W~50.
One is that the shell is a supernova remnant (SNR), and the wings formed due to interactions between the jets and the remnant \citep[e.g.,][]{Goodall_2011a}.
Another is that W~50 was formed only by the backflow of the jets \citep[e.g.,][]{Ohmura_2021}.
In both cases, the wings are attributed to the impact of the jets.
These wings contain X-ray emitting regions known as ``lobes'' \citep[e.g.,][]{Seward_1980,Watson_1983,Yamauchi_1994,Brinkmann_1996}.
Within the lobes, several compact bright structures, referred to as ``knots'' with a size of few-arcmin, are observed \citep[e.g.,][]{Safi-Harb_1997, Safi-Harb_2022,Kayama_2022}, and these knots align with the precession axis of the SS~433 jets. %, suggesting that the jets play a crucial role in the formation of the nebula.

In recent years, very-high-energy (VHE) $\gamma$-ray emissions have been detected using High Altitude Water Cherenkov Observatory \citep[HAWC;][]{Abeysekara_2018}, Large High Altitude Air Shower Observatory \citep[LHASSO;][]{Cao_2024}, and the High Energy Stereoscopic System \citep[H.E.S.S.;][]{{HESS_2024}}.
These observations provide strong evidence for the presence of VHE particles in the jets of SS~433.
The $\gamma$-ray hotspots are located in the eastern and western wings and are spatially coincident with the X-ray lobes. 
The VHE emission extends up to at least 25 TeV, indicating that particles, either electrons or protons, are accelerated to energies exceeding 100~TeV. 
\cite{Abeysekara_2018} concluded that the multi-wavelength spectra, ranging from radio to VHE $\gamma$-rays, can be well explained by leptonic models.
\citet{Sudoh_2020} also showed that the spectrum can be explained by a leptonic model assuming electrons accelerated at the innermost knot.
On the other hand, \citet{Kimura_2020} tested hadronic scenarios and found that hadronic models can also account for the observed multi-wavelength spectra.
Finally, \citet{HESS_2024} reported the discovery of energy-dependent morphology in the eastern and western jets of SS 433, strongly supporting the leptonic origin of the emission.

Hard X-ray emission with a photon index of $\Gamma \sim 1.3-1.6$ is detected from the knots of each lobe \citep[e.g.,][]{Safi-Harb_1999,Moldowan_2005}, and the spectra become softer as one moves away from the innermost knots \citep[e.g.,][]{Safi-Harb_1997,Safi-Harb_1999,Namiki_2000,Brinkmann_2007,Safi-Harb_2022}.
\citet{Safi-Harb_1999} interpreted the X-ray emission as synchrotron emission and attributed the spectral softening to synchrotron cooling. 
Recently, we performed spatially resolved spectroscopy of the western region along the jet precession axis \citep{Kayama_2022}.
Our analysis revealed that the spectral slope gradually steepens along the axis and then rapidly becomes steeper immediately outside the knot w2. 
We also performed model calculations to describe the evolution of the non-thermal particles in the jets.
The comparison between the observation and the model suggests that the particles are transported downstream from the innermost knot.
During this propagation, synchrotron cooling causes the observed spectral evolution.
The knot w2 may correspond to a region where the magnetic field is enhanced, leading to rapid cooling of electrons.
However, spatially resolved X-ray studies of spectral variations, particularly near the knots and in the eastern lobe, remain insufficient to constrain the particle acceleration, transport mechanisms, and emission scenarios.

To obtain a comprehensive understanding of the non-thermal X-ray emission from W~50, we perform new imaging and spectral analyses using XMM-Newton and Chandra.
We perform imaging of almost the entire W~50 and carry out spatially resolved spectral analysis in the eastern region to study its overall structure.
We then discuss particle acceleration in W~50 and SS~433 jets in the context of previous findings from the western side \citep{Kayama_2022}. 
In this study, we assume a distance of 5.5~kpc to W~50/SS~433 \citep[e.g.,][]{Hjellming_1981,Blundell_2004,Lockman_2007}.
%We also adopt an age of W~50 is $2 \times 10^4$~years based on the simulation model of \citet{Goodall_2011a}.
Note that all statistical uncertainties are quoted at the 1$\sigma$ confidence level.

\section{Observations and Data Reduction}
In this work, we utilize nine archival datasets from XMM-Newton and three from Chandra, as listed in table~\ref{tab:imgana:datalist}. 
All datasets were processed following the procedures described below.

\renewcommand{\arraystretch}{1.5}
\begin{table}[htb]
\caption{\centering Observation Data List}
\label{tab:imgana:datalist}
\centering
\begin{minipage}{\textwidth}
\tiny
\begin{tabular}{ccccc}
\hline
Instrument & Obs. ID & Obs. date & Duration [ks] & Region \\ \hline \hline
Chandra &659 & 2000 Jun. 27 & 9.7 & Center \\ 
Chandra &3790 & 2003 Jul. 10 & 58.1 & Center \\ 
Chandra &3843 & 2003 Aug. 21 & 71.2 & West \\ \hline
XMM-Newton & 0137151101 & 2003 Apr. 8 & 17.3 & Center\\
XMM-Newton & 0137150801 & 2003 Oct. 19 & 21.0 & Center\\
XMM-Newton & 0137150901 & 2003 Oct. 25 & 12.0 &  Center\\
XMM-Newton & 0137151801 & 2004 Apr. 15 & 16.0 & Center\\
XMM-Newton & 0075140401 & 2004 Sep. 30 & 32.5 & East\\
XMM-Newton & 0075140501 & 2004 Oct. 4 & 31.3 & East\\
XMM-Newton & 0694870201 & 2012 Oct. 3 & 134.7 & Center\\
XMM-Newton & 0840490101 & 2020 Mar. 24 & 69.1 & East\\
XMM-Newton & 0882560201 & 2021 Oct. 8 & 45.4 & North\\ \hline
\end{tabular}
\end{minipage}
\end{table}
\renewcommand{\arraystretch}{1.0}

The Chandra datasets were reduced using the CIAO 4.12 software package \citep{Fruscione_2006}, with CALibration DataBase (CALDB) version 4.9.8, and the HEADAS software (HEASoft) version 6.29.1. 
Due to the relatively high and non-uniform particle background, as well as the lack of available background data, we excluded the data from the ACIS-S4 chip from our analysis.
For the XMM-Newton datasets, data reduction was performed according to the standard reprocessing and screening criteria described by \citet{Snowden_2014}, using the Science Analysis System (SAS) software version 19.1.0 and the CALDB.

\section{Imaging Analysis}
\subsection{Mosaic Image of Entire W~50}
\label{sec::imag::ana}
To reveal the spatial structure of W~50, we first create a large-area mosaic image.
For the Chandra data, raw images and exposure maps are extracted from the filtered event files using the {\tt fluximage} command in CIAO.
Point-like sources detected using the {\tt wavdetect} software are excluded from the analysis.
The particle background is estimated and subtracted using the ACIS-stowed observations following the method described by \citep{Hickox_2006}.

For the XMM-Newton data, raw images are generated using the {\tt mos-spectra} and {\tt pn-spectra} tasks in the SAS software.
Point sources are identified and removed using the {\tt cheese} task, which detects point sources in full-field images and creates corresponding mask files. 
Model background images for the quiescent particle background (QPB) and soft protons (SP) are generated following the methods described by \citet{Snowden_2014}.
The SP component parameters will be estimated in sub-section~\ref{sec:spec_analysis}.

We then generate background-subtracted images and exposure maps for each observation in two energy bands, 0.5--1.5 keV (low) and 2.0--7.0 keV (high).
Finally, we combine these individual images into single mosaic images for each energy band using the {\tt ximage} software in HEASoft. 

\subsection{Morphology of X-ray Extended Jet of SS~433}
\label{sec:img_morph}

\begin{figure}[htbp]
\includegraphics[width=0.5\textwidth]{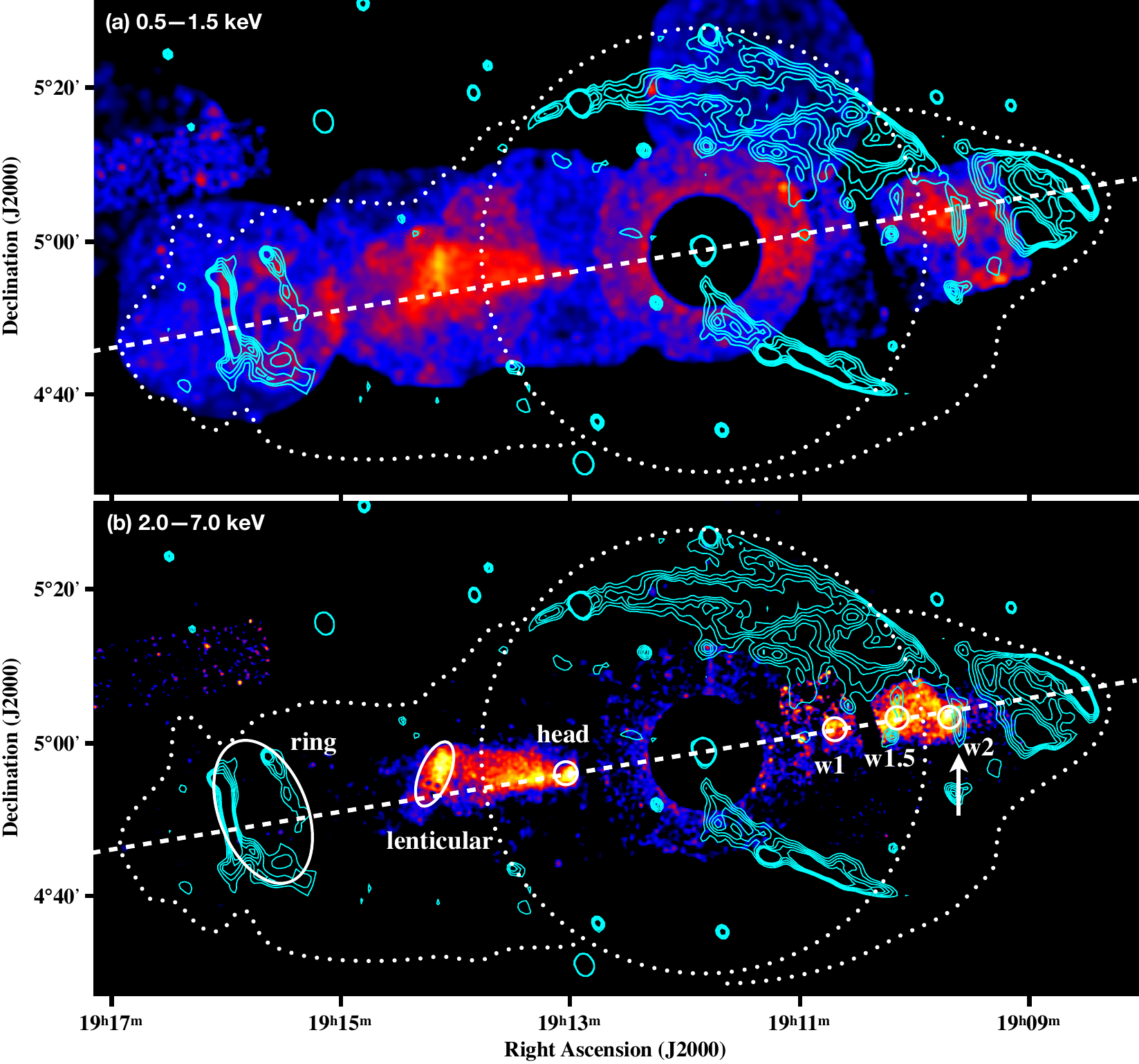}
\caption{X-ray image around W~50 in the energy band of (a) 0.5--1.5~keV and (b) 2.0--7.0~keV. The white dashed line corresponds to the jet precession axis determined by \citet{Eikenberry_2001}. The dotted lines and cyan contours indicate approximately enclosed regions measured from the flux densities and the 140 MHz LOFAR radio continuum, respectively \citep{Broderick_2018}.
The locations of the knots are marked with white circles. The white arrow indicates ”filament J” as identified by \citet{Goodall_2011b}. {Alt text: This figure shows the X-ray image of the entire W~50 nebula, composed of two panels.}}
\label{fig:ent_img_el_a}
\end{figure}

Figure~\ref{fig:ent_img_el_a} shows the X-ray mosaic image of the entire W~50.
A circular region with a 7\farcm5 radius around SS~433 is excluded to enhance the visibility of faint diffuse structures.
The soft X-ray distribution appears to be more extended, while the hard X-rays are visible only around the jet precession axis of SS~433.
Several bright structures, known as knots, are found: the head (e1), lenticular (e2), and ring (e3) in the eastern lobe, and w1, w1.5, and w2 in the western lobe.
The distances from SS~433 to the head and w1 are measured to be $\ang{;18;}$ and $\ang{;16;}$, corresponding to 29~pc and 26~pc, respectively.
The knots "head" and "lenticular" were originally defined by \citet{Brinkmann_2007} and \citet{Safi-Harb_2022}.
The definitions of knots w1 and w2 differ from those by \citet{Safi-Harb_1997} and have been redefined by \citet{Kayama_2022}.

\begin{figure}[htbp]
\includegraphics[width=0.5\textwidth]{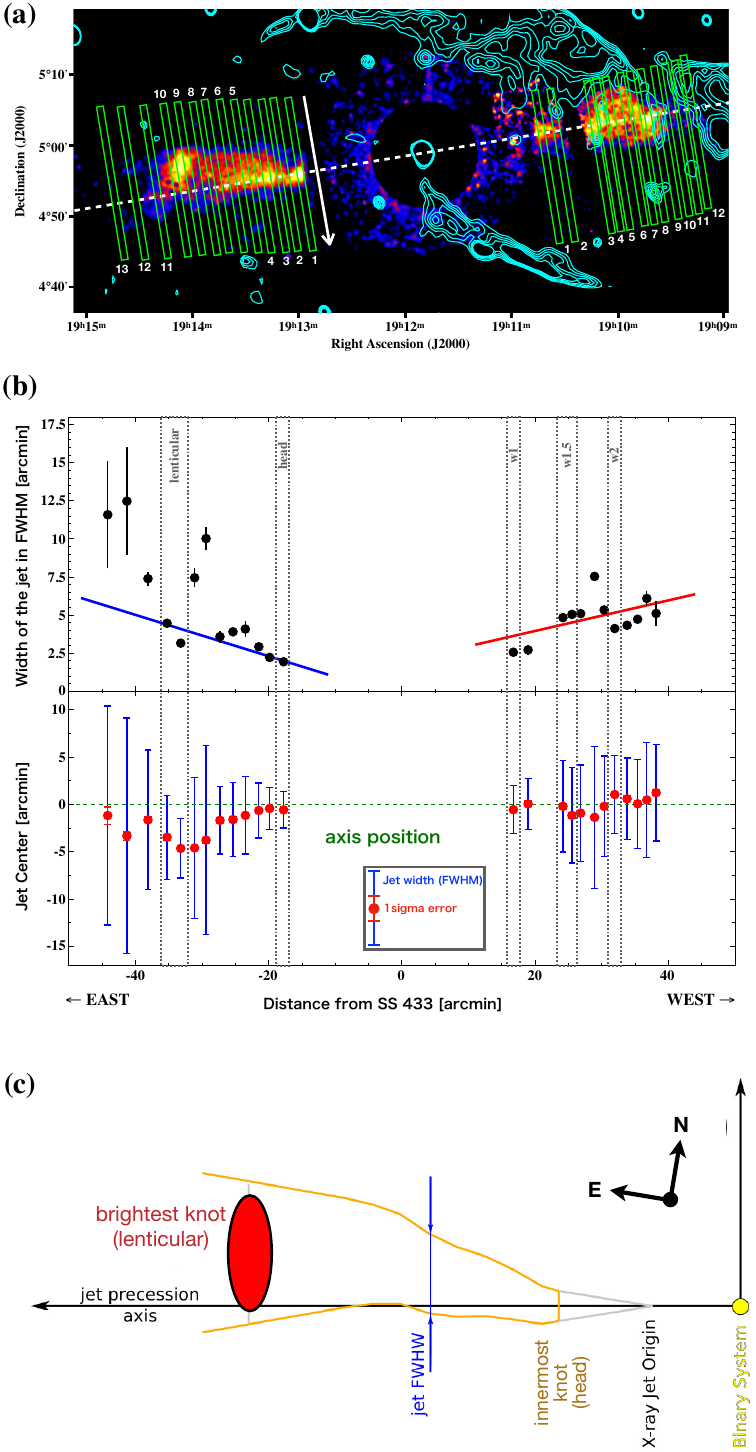}
\caption{(a) X-ray image in the energy band of 2.0--7.0 keV. The rectangles represent the regions used for the jet opening angle measurement. The dashed line indicates the jet precession axis. (b; Top) Widths of the jets in FWHM as a function of angular distance from SS~433. The blue and red lines show the result of linear regression. (b; Bottom) The peak intensity position and jet width as a function of angular distance from SS~433. The green line indicates the position of the jet precession axis. (c) A schematic view of the SS~433 eastern jet inferred from our analysis results. {Alt text: This figure consists of three panels. The upper panel (a) shows an X-ray image of the W~50 nebula. The middle panel (b) shows two line graphs: one showing the profile of jet width and the other showing a plot of the peak intensity position. The lower panel (c) shows a schematic drawing of the eastern jet.}}
\label{fig:ent_img_el_c}
\end{figure}

%Our previous study showed that the X-ray emission in the high-energy band is dominant along the jet precession axis in the western lobe \citep{Kayama_2022}.
In our previous study, we quantified the geometry of the western jet using high-energy band X-ray images \citep{Kayama_2022}.
We inferred that the shape of the western jet is conical with a half-opening angle of $2\overset{\circ}{.}6 \pm 0\overset{\circ}{.}3$, based on the observed width of emitting regions and their distances from SS~433.
In order to quantify the geometry of the eastern jets, we measure the jet width using the high-energy band image shown in figure~\ref{fig:ent_img_el_a}.
The opening widths of the jets are estimated from the profiles in each rectangular region in Figure~\ref{fig:ent_img_el_c}(a).
According to \citet{Kayama_2022}, we fit the profiles with a phenomenological model consisting of a Gaussian and a constant component, which represent the W~50-originated emission from the jet and the background emission, respectively. 
We set the intensity of both components and the width of the Gaussians as free parameters.
As seen in figure~\ref{fig:ent_img_el_a}, the eastern jet appears to bend northward, rather than extending straight like the western jet.
Therefore, the center positions of the Gaussians are also set as free parameters.

Figure~\ref{fig:ent_img_el_c}(b) shows the obtained widths in FWHM plotted against the angular distance from SS~433, respectively.
Note that the background contribution has been subtracted.
Based on the increase in width with distance, the jet shape is considered to be conical, similar to the western jet.
Fitting a linear model to the data points in figure~\ref{fig:ent_img_el_c}(b) gives a half-opening angle of $3\overset{\circ}{.}9 \pm 0\overset{\circ}{.}2$, and our results also indicate that the eastern jet is bending northward.
%Therefore, based on our measurements, the morphology of the eastern jet is inferred to be as shown in figure~\ref{fig:ent_img_el_c}(c).
A sketch illustrating the procedure of inferring the jet parameters is shown in figure~\ref{fig:ent_img_el_c}(c).
Note that if the X-ray jets are formed by recollimation of the precessing jets at smaller distances from SS~433, as suggested by \citet{Monceau-Baroux_2015}, the linear extrapolation to the jet origin does not necessarily match the position of the binary system.
We therefore treat the jet origin as a free parameter.

\section{Spectral Analysis}
\subsection{Global Spectra of Eastern Lobe and Northern Part of W~50}
\label{sec:spec_analysis}

\begin{figure}[htbp]
\includegraphics[width=0.5\textwidth]{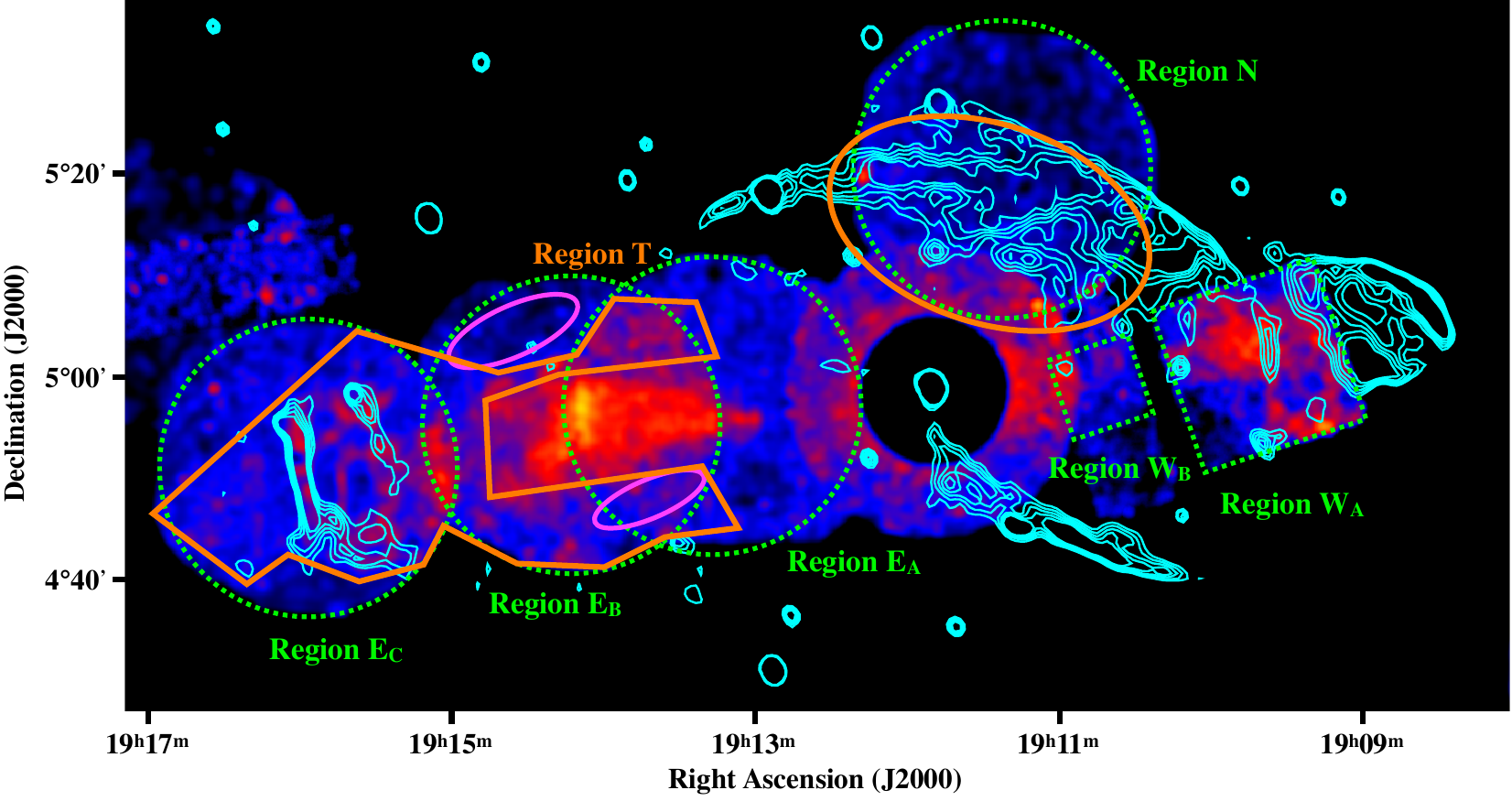}
\caption{X-ray image of W~50 in the energy band of 0.5--1.5 keV, the same as figure 1(a), with the regions from which spectra were extracted for our spectral analysis indicated by green dashed lines. 
The region enclosed by the orange lines indicates the thermal emission-dominated region (region T).
The magenta ellipses indicate the faint regions used for background estimation. {Alt text: This figure shows the X-ray image of the entire W 50 nebula with the regions from which spectra were extracted for our spectral analysis.}}
\label{fig:spec_extreg_glob_a}
\end{figure}

We first analyze spectra spatially integrated over the entire fields of view (FOVs) of each observation of the eastern lobe and the northern part of the W~50 central shell.
As shown in figure~\ref{fig:spec_extreg_glob_a}, the regions are defined according to the FOVs of each observation.
Note that the X-ray emission from regions $\rm W_A$ and $\rm W_B$ (corresponding to regions A and B in \citealt{Kayama_2022}) was studied in our previous work.
Here, we mainly focus on regions $\rm E_A$, $\rm E_B$, $\rm E_C$, and N.
We extract spectra and response files following the methods described by \citet{Snowden_2014}, using the SAS software.

\begin{figure*}[htbp]
\begin{center}
\includegraphics[width=0.95\textwidth]{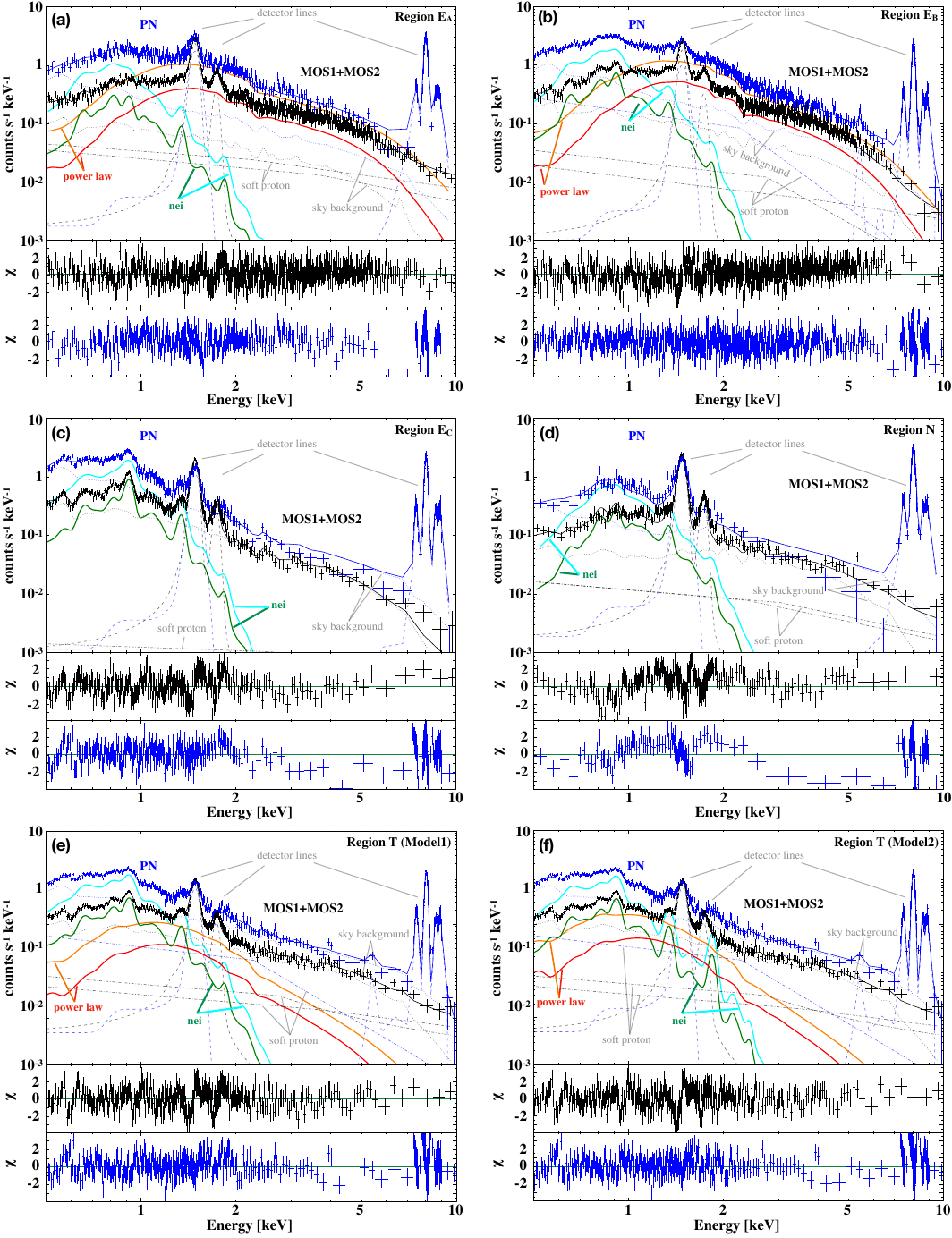}
\end{center}
\caption{Spectra extracted from (a) Region $\rm E_A$, (b) Region $\rm E_B$, (c) Region $\rm E_C$, (d) Region N, and (e)--(f) Region T with the best-fit curves. 
Note that the model curves in (e) and (f) show the best-fit results for Model 1 and Model 2, respectively.
The black and blue points correspond to the data from MOS1+MOS2 and PN detectors. The non-thermal (power-law) model is shown by the red (for MOS1+MOS2 data) and orange (for PN data) lines, and the thermal (NEI) model is shown by the green (for MOS1+MOS2 data) and cyan (for PN data) lines.
The thick-solid lines indicate the total source emission models. 
The thin dot, dashed dot, and dashed lines indicate the sky background, SP, and line components of the background, respectively. 
The bottom panels present residuals from the best-fitting models. {Alt text: This figure consists of six panels, each showing multi-line graphs representing the X-ray spectra and fitting results.}}
\label{fig:xmmspec1}
\end{figure*}

Figure~\ref{fig:xmmspec1}(a)--(d) shows the spectra obtained from regions $\rm E_A$, $\rm E_B$, $\rm E_C$, and N.
These spectra consist of W~50-originated emission, the Galactic and extragalactic X-ray background, and a background component originating from high-energy particles.
Since the W~50 nebula extends beyond the FOV, background spectra cannot be directly extracted from these datasets.
Therefore, we adopt the same sky-background model as described by \citet{Kayama_2022} for our study.
Here, we assume contributions from the Galactic ridge X-ray emission (GRXE), the cosmic X-ray background, and foreground emission \citep[FE; e.g.,][]{Uchiyama_2013}.
Note that the normalization of the GRXE depends on the Galactic coordinates, and is set to the value at coordinates ($l$, $b$) = ($39.92$, $-2.80$) for Region~$\rm E_A$, $\rm E_B$, $\rm E_C$ and ($l$, $b$) = ($39.69$, $-2.24$) for Region N, respectively.

Since the spectrum of the particle background depends on the detector, we cannot apply the same model described by \citet{Kayama_2022}.
The XMM-Newton EPIC particle background consists of two components, QPB and SP, which are estimated according to \citet{Snowden_2014}.
We generate the model QPB with the {\tt mos-back} and {\tt pn-back} for the MOS and PN detectors, respectively, and subtract these from the spectra.
To remove the SP component, we use the {\tt mos-filter} and {\tt pn-filter} scripts to exclude the SP flaring time intervals.
However, the screening task cannot remove all soft-proton contamination \citep[e.g.,][]{Henley_2013,Snowden_2014}, so we model it with a broken power law unmodified by the instrumental response.

Since the intensity of the SP component is low and cannot constrain the photon indices and break energy, we estimate the parameters of the background spectra as follows. 
We assume that the spectral shape of the SP component is the same during both the flaring period ("flare") and when the flare rate is low ("clean"). 
The SP spectra are extracted by subtracting the clean spectrum from the flare spectrum (flare - clean).
We then fit these spectra with the SP model to constrain the photon indices and cutoff energy.
Instrumental line components (Al, Si, Ni, Cu, Zn) are modeled using Gaussians with fixed energies and zero width, following \citep{Snowden_2014}.
Note that for Region $\rm E_B$, the flare duration is short, making it impossible to constrain the SP model parameters using this method.
Instead, we extract spectra from the faint region, marked by magenta ellipses in figure~\ref{fig:spec_extreg_glob_a}, and fit them simultaneously with the W50-originated emission and background model.

\renewcommand{\arraystretch}{1.5}
\begin{table*}
\caption{\centering Best-Fit Parameters of W~50-Originated Components for Regions~$\rm E_A$, $\rm E_B$, $\rm E_C$, and N}
\begin{threeparttable}[tbp]
\footnotesize
\begin{tabularx}{\textwidth}{cccccc}
\hline
 \textbf{Model function}&  \textbf{Parameter}& \multicolumn{4}{c}{\textbf{Value}} \\
&&\textbf{Region~$\rm E_A$}&\textbf{Region~$\rm E_B$}&\textbf{Region~$\rm E_C$}&\textbf{Region~N}\\
 \hline
TBabs&$N_\mathrm{H}$ ($\mathrm{10^{22}~cm^{-2}}$)& $0.91^{+0.05}_{-0.02}$ & $1.09\pm0.03$ & $1.17\pm0.01$ & $1.74\pm{0.02}$\\
power law&$\Gamma$& $1.99^{+0.02}_{-0.03}$ & $2.47\pm0.03$ & $-$ & $-$ \\
 &norm \tnote{a}& $3.87^{+0.09}_{-0.07}\times10^{-3}$ & $7.17^{+0.19}_{-0.17}\times10^{-3}$ & $-$ & $-$ \\
 vnei& $kT_{\mathrm{e}}$ (keV)& $0.26\pm0.01$ & $0.21\pm0.00$ & $0.19\pm0.00$ & $0.25^{+0.01}_{-0.00}$\\
    & elemental abundances (solar)& \multicolumn{4}{c}{$1$ (fixed)}\\ 
  & $\tau$ ($\mathrm{10^{11}~s~cm^{-3}}$ ) & $\leq 310$ & $\leq 19.29$ & $2.04^{+0.19}_{-0.18}$ & $\leq 413$ \\
    & VEM\tnote{b} ($\mathrm{cm^{-3}}$)& $4.73^{+0.85}_{-0.45}\times 10^{57}$ & $3.19^{+0.30}_{-0.32}\times 10^{58}$ & $8.65^{+0.25}_{-0.13}\times 10^{58}$ & $2.15^{+0.33}_{-0.66}\times 10^{58}$ \\
    \\
&$\chi^2$/d.o.f \tnote{c}& $5150.91/4623$ & $4851.82/4495$ & $3862.50/3461$ & $3441.20/3143$\\
\hline
\end{tabularx}
\label{tab:spec:xmm_east}
\begin{tablenotes}
\footnotesize
\item[a]{The unit is $\mathrm{photons~s^{-1}~cm^{-2}~keV^{-1}~at~1~keV}$.}
\item[b]{Volume emission measure (VEM) is defined as $\int n_\mathrm{e}n_\mathrm{H}dV$, where $n_{e}$, $n_\mathrm{H}$, and $V$ are the electron density, hydrogen density, and emitting volume, respectively.}
\item[c]{d.o.f.: degree of freedom.}
\end{tablenotes}
\end{threeparttable}
\end{table*}
\renewcommand{\arraystretch}{1.0}

Several previous studies \citep[e.g.,][]{Brinkmann_2007,Safi-Harb_2022, Kayama_2022} have reported that the W~50-originated emission component can be described by a superposition of a power-law model and a non-equilibrium ionization (NEI; {\tt vnei} model in XSPEC) plasma model, both of which are modified by interstellar absorption.
Here, we assume the same model for the W~50-originated emission.
We adopt the Tuebingen-Boulder absorption model \citep[{\tt tbabs};][]{Wilms_2000} for the interstellar absorption.
The column density $N_{\rm H}$, photon index $\Gamma$, electron temperature $kT_{\rm e}$, ionization timescale $\tau$, and these normalizations are set as free parameters.
Note that the elemental abundances are fixed to the solar values due to limited statistics. 
The best-fit model curve and parameters of W~50-originated emissions are shown in figure~\ref{fig:xmmspec1}(a)--(d) and table~\ref{tab:spec:xmm_east}, respectively.
We detect the non-thermal emission from regions $\rm E_A$ and $\rm E_B$ with a significance level $\gg 10\sigma$, but not from Region $\rm E_C$ and N ($< 2\sigma$).
%The slopes of the non-thermal emission become steeper with the distance from SS~433, a trend that is consistent with previous studies \citep[e.g.,][]{Safi-Harb_2022}.
The spectrum of the non-thermal emission becomes steeper with increasing distance from the jet origin, a trend that is consistent with previous studies \citep[e.g.,][]{Safi-Harb_2022}.
The parameters of thermal emission are also roughly consistent with previous studies \citep[e.g.,][]{Brinkmann_2007, Safi-Harb_2022} and also comparable to those in the western lobe \citep[e.g.,][]{Kayama_2022}.

In order to examine the parameters of the thermal emission, we next extract the spectrum from the regions in which the X-ray emission is dominated by the thermal component.
The region from which the spectrum is extracted is indicated in figure~\ref{fig:spec_extreg_glob_a}, and the spectrum is shown in figure~\ref{fig:xmmspec1}(e)--(f).
We fit the spectrum with the same model as described above, setting $N_{\rm H}$, $\Gamma$, $kT_{\rm e}$, $\tau$, and the normalizations as free parameters, while the elemental abundances are fixed to the solar values (Model~1).
%The best-fit parameters are listed in table~\ref{tab:spec:thermal_polygon}, which are roughly consistent with previous studies in the outer part of the eastern lobe \citep[e.g.,][]{Brinkmann_2007,Safi-Harb_2022} and the western lobe \citep[e.g.,][]{Kayama_2022}.
The best-fit model and these parameters are shown in figure~\ref{fig:xmmspec1}(e) and table~\ref{tab:spec:thermal_polygon}, respectively.
These results are roughly consistent with previous studies on the outer part of the eastern lobe \citep[e.g.,][]{Brinkmann_2007,Safi-Harb_2022} and the western lobe \citep[e.g.,][]{Kayama_2022}.
Since excess emissions are found around the energies of lines originating from highly ionized elements such as O, Ne, Mg, and Fe, we next set the elemental abundances of these elements as free parameters.
The best-fit model and these parameters are shown in figure~\ref{fig:xmmspec1}(f) and table~\ref{tab:spec:thermal_polygon}, respectively (Model 2).
Based on these two results, we perform an F-test using the {\tt ftest} tool in XSPEC and find that the fitting is significantly improved with a significance level $> 7\sigma$.
The parameters measured in the eastern lobe and the northern part of the central shell are consistent with those in the western lobe, implying that the entire nebula is uniformly filled with thermal plasma, which may have a common origin.

\renewcommand{\arraystretch}{1.8}
\begin{table}[h]\centering
\centering
\begin{minipage}{1\textwidth}
\begin{threeparttable}[b]
\caption{Best-Fit Parameters of W~50-Originated Thermal and non-thermal \\emission.}
\tiny
\begin{tabularx}{0.5\textwidth}{cccc}
\hline
\hline
&&\textbf{Model~1}&\textbf{Model~2} \\
Model function& Parameter&& \\
\hline
TBabs&N$_\mathrm{H}$ $(10^{22}~\mathrm{cm}^{-2})$&$\mathrm{1.12}^{+0.02}_{-0.02}$&$0.78^{+0.04}_{-0.04}$\\
pwgpwr&$\Gamma$&$4.04^{+0.12}_{-0.17}$&$2.93^{+0.32}_{-0.25}$\\
&flux \tnote{a}&$0.97^{+0.05}_{-0.08}$&$0.47^{+0.07}_{-0.06}$\\
vnei&$kT_{\mathrm{e}}$ ($\mathrm{keV}$)&$0.21^{+0.00}_{-0.01}$&$0.30^{+0.02}_{-0.01}$\\
&Z$_\mathrm{O}$ (solar)&$1$(fix)&$0.28^{+0.07}_{-0.04}$\\
&Z$_\mathrm{Ne}$ (solar)&$1$(fix)&$0.54^{+0.11}_{-0.07}$\\
&Z$_\mathrm{Mg}$ (solar)&$1$(fix)&$0.57^{+0.10}_{-0.06}$\\
&Z$_\mathrm{Fe}$=Z$_\mathrm{Ni}$ (solar)&$1$(fix)&$0.23^{+0.04}_{-0.03}$\\
&$\tau~(\mathrm{s~cm^{-3}}$)&$1.63^{+0.61}_{-0.15}$$\times\mathrm10^{11}$&$2.58^{+0.62}_{-0.85}$$\times\mathrm10^{11}$\\
&VEM \tnote{b} ($\mathrm{cm}^{-3}$)&$2.05^{+0.29}_{-0.08}$$\times\mathrm10^{58}$& $6.84^{+0.01}_{-0.19}$$\times\mathrm10^{57}$ \\
\hline
&$\chi^2$/d.o.f. \tnote{c}&$3135.84/3742$&$3081.44/3738$\\
\hline
\end{tabularx}
\label{tab:spec:thermal_polygon}
\begin{tablenotes}\tiny
\item[a]{The normalizations correspond to the flux in unit of $10^{-12}~\mathrm{erg~cm^{-2}~s^{-1}}$, which is integrated over the energy range of $1.0$--$7.0$~keV in this analysis.}
\item[b]{Volume emission measure (VEM) is defined as $\int n_\mathrm{e}n_\mathrm{H}dV$, where $n_{e}$, $n_\mathrm{H}$, and $V$ are the electron density, hydrogen density, and emitting volume, respectively.}
\item[c]{d.o.f: degree of freedom.}
\end{tablenotes}
\end{threeparttable}
\end{minipage}
\end{table}
\renewcommand{\arraystretch}{1}

\subsection{Spectral Variation Along the Jet Precession Axis}
\begin{figure}[htbp]
\includegraphics[width=0.5\textwidth]{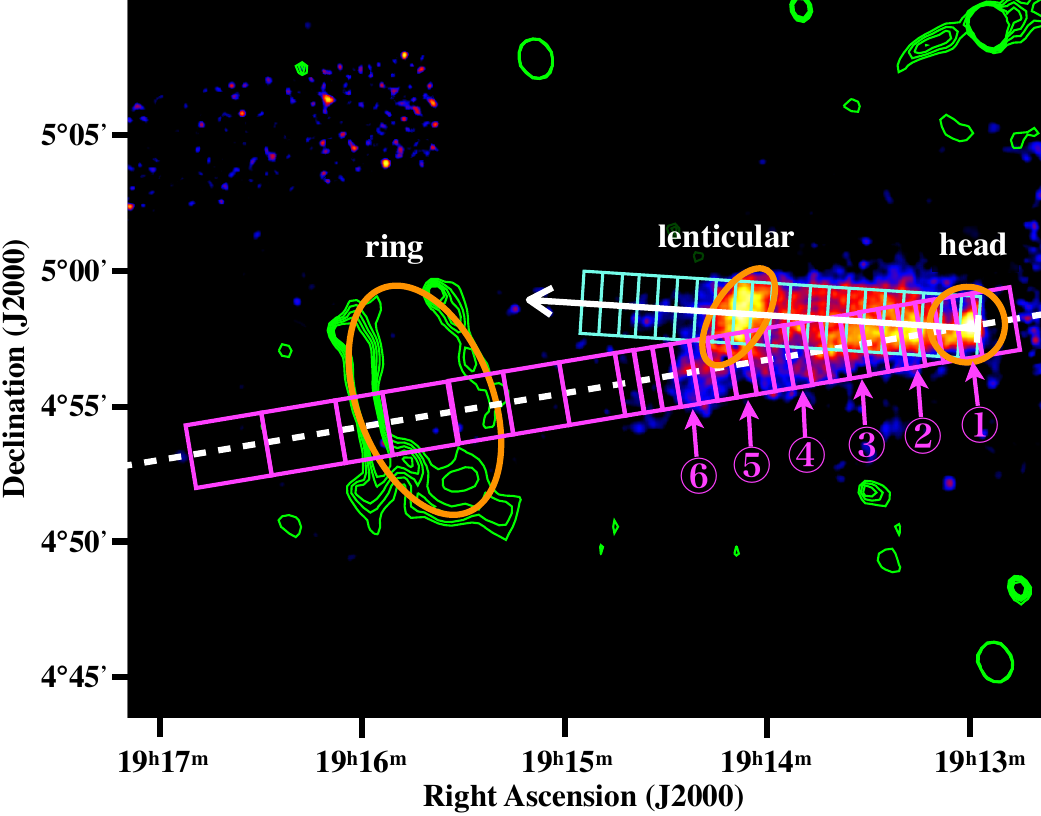}
\caption{X-ray image in the energy band of 2.0--7.0 keV. The magenta and cyan boxes indicate the spectra extraction regions for the analysis in Figure~\ref{fig:xc_onaxis_profile} and Figure~\ref{fig:dist_knotaxis}, respectively. The knot locations are indicated by the orange ellipses. The green contours indicate the 140~MHz LOFAR radio continuum \citep{Broderick_2018}. The magenta numbers indicate the spectra extraction regions shown in figure~\ref{fig:spec_onaxis}. {Alt text: This figure shows the X-ray image of the eastern region of the W~50 nebula.}}
\label{fig:xc_onaxis}
\end{figure}

\begin{figure*}[htbp]
\includegraphics[width=\textwidth]{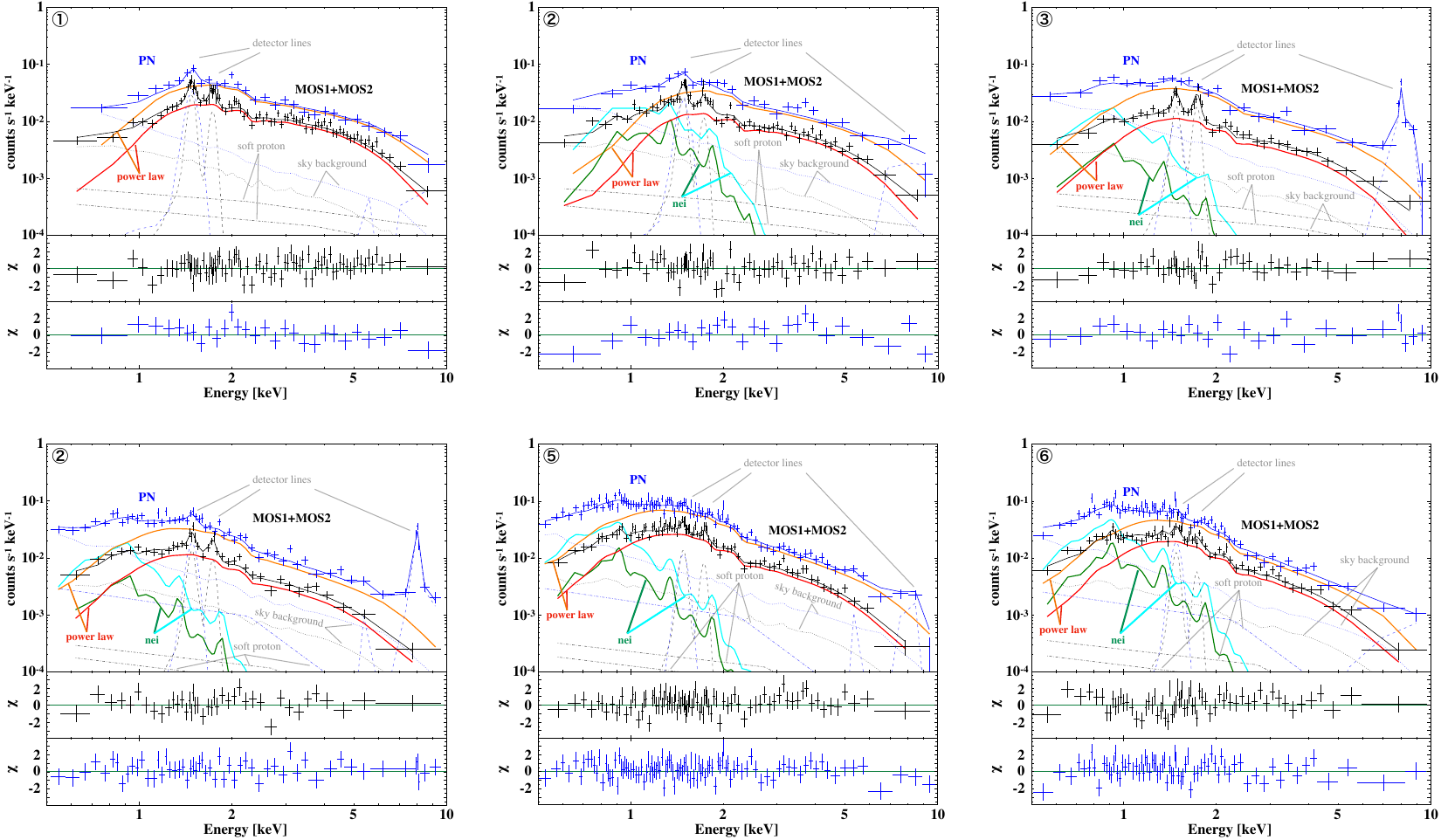}
\caption{X-ray spectra extracted from the regions indicated in figure~\ref{fig:xc_onaxis}. The colors and line types are the same as in Figure~\ref{fig:xmmspec1}. {Alt text: This figure consists of six panels, each showing multi-line graphs representing the X-ray spectra and fitting results. }}
\label{fig:spec_onaxis}
\end{figure*}

As shown by several previous studies \citep[e.g.,][]{Yamauchi_1994,Safi-Harb_1999,Namiki_2000,Brinkmann_2007,Kayama_2022}, the non-thermal emission tends to have a softer spectrum as one moves away from SS~433 along the jet precession axis. 
Our analysis presented in sub-section~\ref{sec:spec_analysis} also indicates a similar trend in the eastern lobe, where the spectrum in Region $\rm E_B$ is softer than that in Region $\rm E_A$.
To obtain a more detailed view of the spectral variation of the non-thermal emission along the axis, we perform spatially resolved spectroscopy.
Spectra are extracted from the regions shown in figure~\ref{fig:xc_onaxis}, and representative X-ray spectra from some of these regions are shown in figure~\ref{fig:spec_onaxis}.
As shown in figure~\ref{fig:xmmspec1}, the X-ray emission along the jet precession axis requires both non-thermal and thermal components.
Based on the results in sub-section~\ref{sec:spec_analysis}, the parameters of the hot plasma are nearly uniform across the nebula.
Therefore, due to limited statistics and the dominance of non-thermal emission, we assume that the thermal emission is uniform in the eastern lobe.
Accordingly, the parameters of the thermal component are fixed to those of Model 2 listed in table~\ref{tab:spec:thermal_polygon}, with only the normalization left as a free parameter.

\begin{figure}[htbp]
\includegraphics[width=0.5\textwidth]{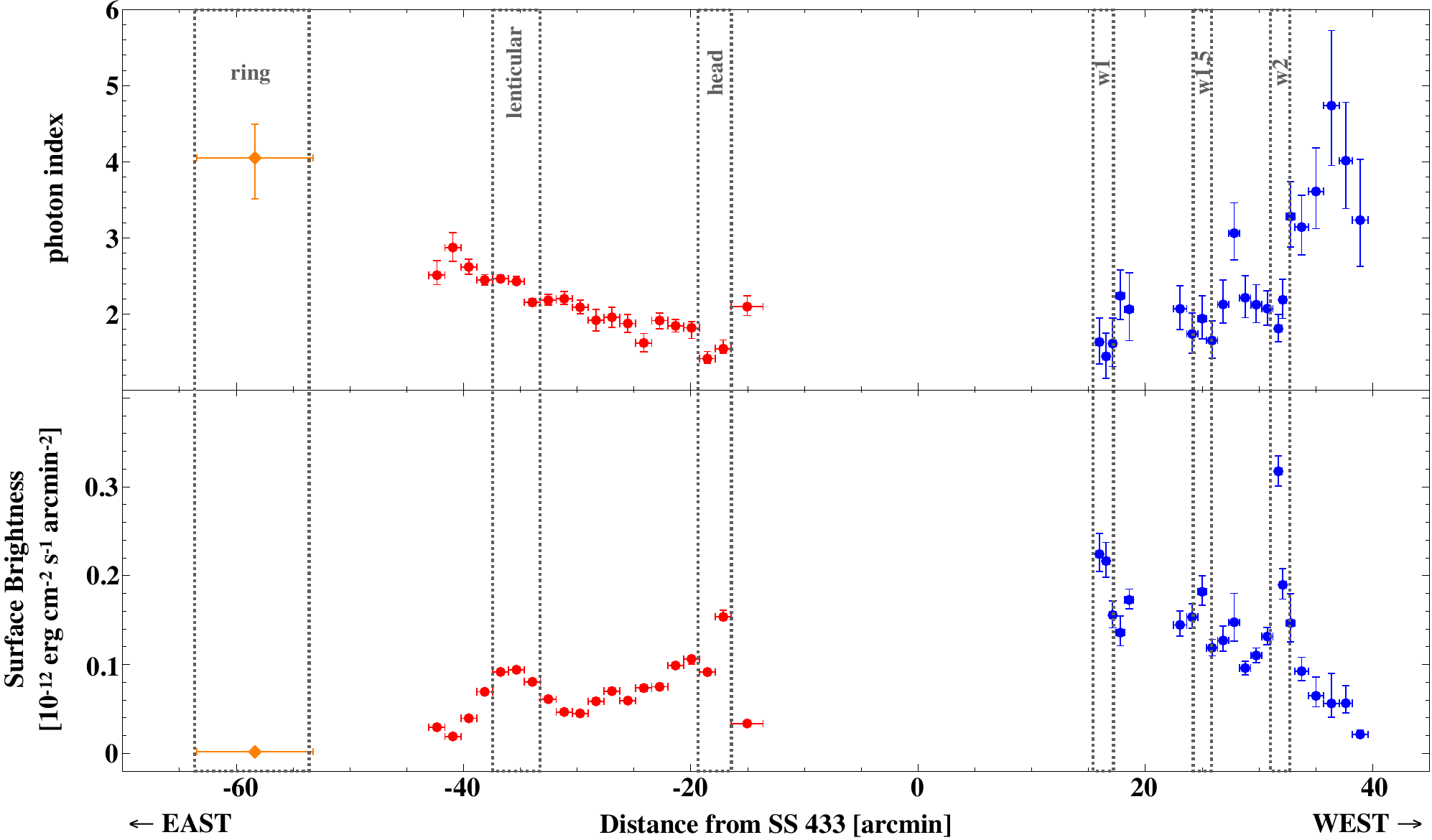}
\caption{Photon index (top panel) and surface brightness in 1.0--7.0 keV (bottom panel) plotted against the distance from SS~433. The red and blue data points correspond to the eastern and western jets, respectively. Note that the data points for the western lobe are referred from \citet{Kayama_2022}. The orange data points correspond to the ring filament. The regions between the dashed lines correspond to the X-ray bright knots. {Alt text: Graphs representing the profile of the fitting results of the power-law model.}}
\label{fig:xc_onaxis_profile}
\end{figure}

Figure~\ref{fig:xc_onaxis_profile} shows the spatial variations of the photon index and surface brightness of the non-thermal component, plotted as functions of distance from SS~433.
Note that the data points for the western lobe are taken from \citet{Kayama_2022}.
The values at the "ring" region are also included in the figure.
The non-thermal emission shows spectral steepening as one moves away from SS~433. 
We find no significant non-thermal X-ray emission in the regions between SS~433 and the head.

\begin{figure}[htbp]
\includegraphics[width=0.5\textwidth]{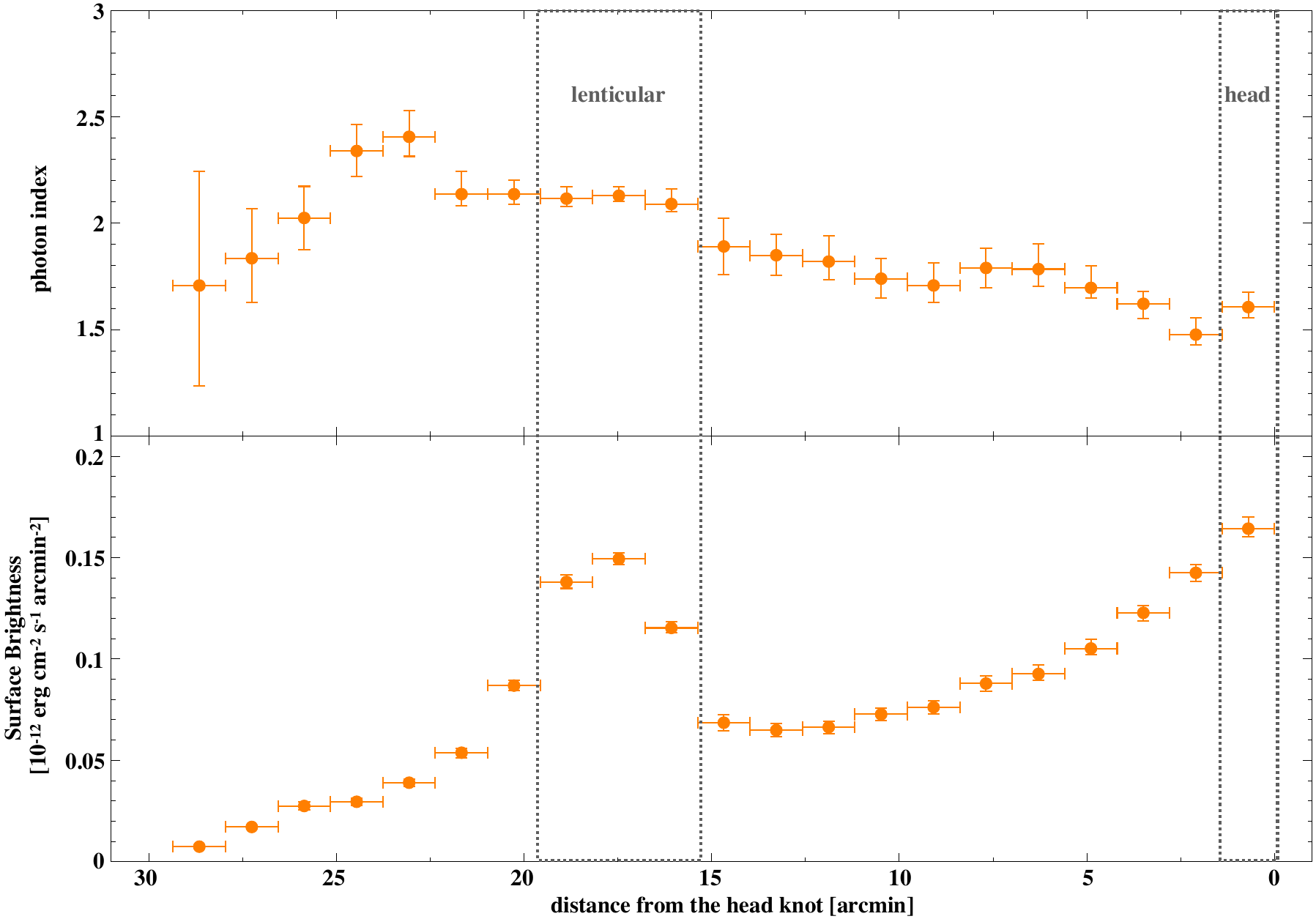}
\caption{Photon index (top panel) and surface brightness variations in 1.0--7.0 keV (bottom panel) along the axis defined in Figure~\ref{fig:xc_onaxis} with the cyan boxes. The origin of the horizontal axis is the location of the head. The regions between the dashed lines correspond to the X-ray bright knots. {Alt text: Graphs representing the profile of the fitting results of the power-law model.}}
\label{fig:dist_knotaxis}
\end{figure}

As discussed in sub-section~\ref{sec:img_morph}, the western jet propagates along the precession axis, whereas the eastern jet does not appear to be perfectly aligned with it.
If this is the case, the spectral variation along the precession axis and along the jet itself may differ.
Therefore, as shown in figure~\ref{fig:xc_onaxis}, we redefined the axis in the direction from the head to the lenticular region and examined the spectral variations along this axis.
We apply the same background model and W~50-originated emission model as in the previous spectral fitting.
Figure~\ref{fig:dist_knotaxis} presents the profiles of the spectral parameters along the axis defined in Figure~\ref{fig:xc_onaxis}.
We find that in the western lobe, the spectra rapidly steepen immediately outside knot w2 \citep{Kayama_2022}, whereas in the eastern lobe, the spectra gradually steepen from the head knot toward the east.

\section{Discussion}

\subsection{Non-thermal Emission and its Variation}
Our spectral analysis indicates that a non-thermal component dominates the X-ray emission in both the eastern and western lobes. 
The best-fit value of $\Gamma$ for the head is roughly consistent with previous studies \citep[e.g.,][]{Safi-Harb_2022} and is in good agreement with that of the innermost knot of the western lobe \citep[knot w1;][]{Kayama_2022}. 
These results suggest that the non-thermal emission in the eastern lobe is also synchrotron emission, as in the western lobe. 
Here, we assume that the X-ray emission from both lobes has a common origin, as proposed by \citet{Sudoh_2020} and \citet{Kayama_2022}.
In this scenario, electrons are accelerated to relativistic energies at the innermost knots and lose energy via synchrotron cooling as they propagate along the jet.
The magnetic fields in regions such as w2 and the lenticular are locally enhanced, given that synchrotron emissivity is highly sensitive to the magnetic field strength.

To quantitatively reproduce the spectral variation of the synchrotron emission in the eastern lobe, we apply the same model as described in \citet{Kayama_2022}.
The energy-spatial density distribution of electrons at the injection site is assumed to follow a power-law with an exponential cutoff.
The index of the injection spectrum is set to $p_{\rm inj}$ = 2.08, corresponding to a synchrotron photon index of $\Gamma$ = 1.54 in the slow cooling regime, which was obtained from the head knot in our spectral analysis.
We assume the cutoff energy of the injection spectrum to be $E_{\rm cut}$ = 1.5~PeV, as derived from multi-wavelength spectral modeling by \citet{Sudoh_2020}.
We examine three distinct values for the jet velocity ($v_{\rm jet}$): (i) the jet maintains its initial velocity ($v_{\rm jet}$ = 0.26$c$), (ii) $v_{\rm jet}$ = 0.10$c$ as estimated by \citet{Panferov_2017} for the head knot, and (iii) $v_{\rm jet}$ = 0.26$c$/4 = 0.065$c$, based on the assumption of a strong shock occurring at the knot.
We examine three different cases for the jet velocity ($v_{\rm jet}$): (i) the jet maintains its initial velocity ($v_{\rm jet}$ = 0.26$c$), (ii) $v_{\rm jet}$ = 0.10$c$, as estimated for the head knot by \citet{Panferov_2017}, and (iii) $v_{\rm jet}$ = 0.26$c$/4 = 0.065$c$, assuming a strong shock at the knot.
This assumption is the same as that for the western side \citep{Kayama_2022}.

\begin{figure*}[htbp]
\includegraphics[width=\textwidth]{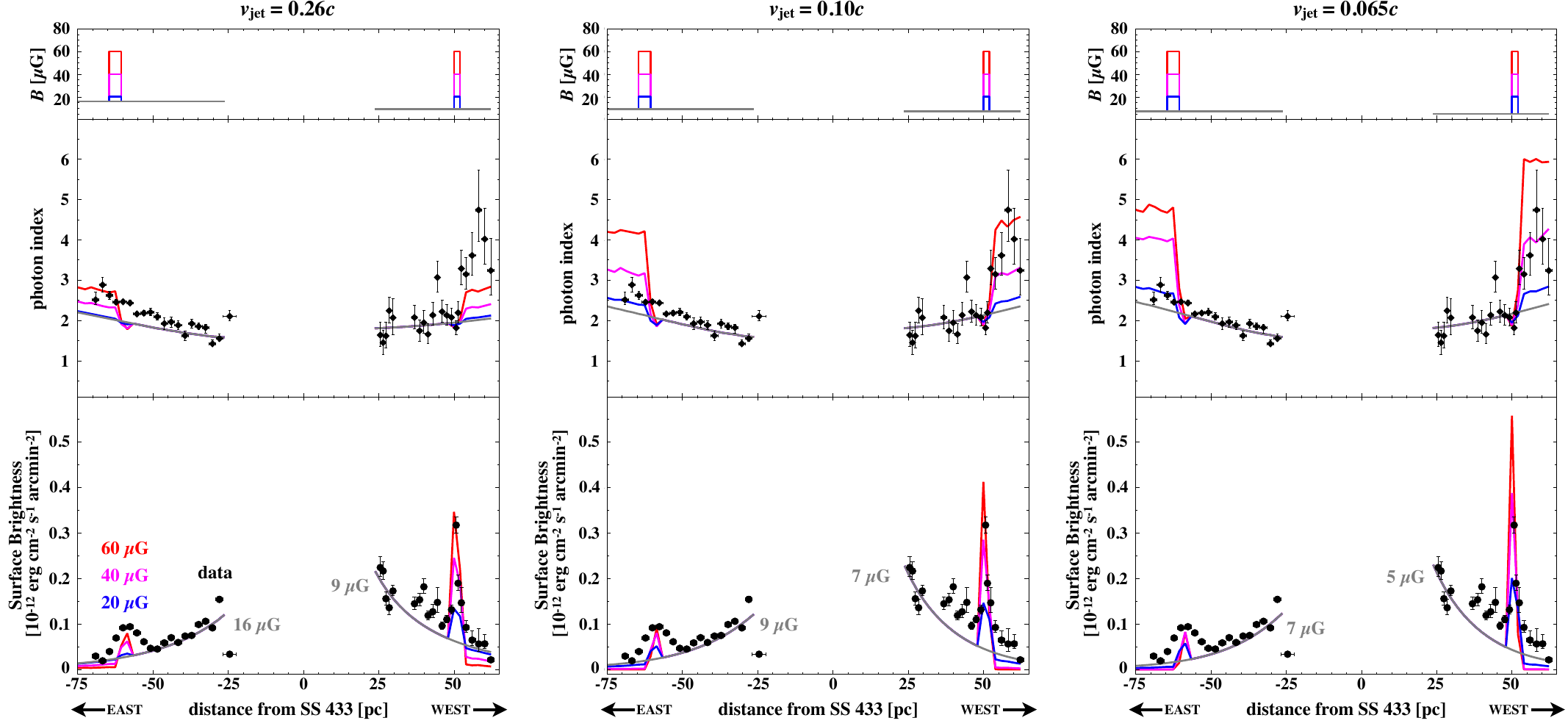}
\caption{Comparison between the model and the observed spectral variation along the jet precession axis. The results for the western side are taken from \citet{Kayama_2022}. The top panels show magnetic field strength profiles assumed in the model. The middle and bottom panels present the photon index and surface brightness profiles, respectively. The gray curves represent models with uniform magnetic field strengths ($16~\mu G$, $9~\mu G$, and $7~\mu G$ for the eastern side, and $9~\mu G$, $7~\mu G$, and $5~\mu G$ for the western side, respectively). The blue, magenta, and red curves represent profiles at the lenticular with $20~\mu G$, $40~\mu G$, and $60~\mu G$, respectively. The left-hand, center, and right-hand panels show model curves with $v_{\rm jet}$ = 0.26$c$, 0.10$c$, and 0.065$c$, respectively. {Alt text: This figure consists of three groups of three-panel, multi-line graphs.}}
\label{fig:model_east}
\end{figure*}

Figure~\ref{fig:model_east} shows the model calculation results overlaid with the observational data for both the eastern jet (from figure~\ref{fig:xc_onaxis_profile}) and the western jet \citep[from][]{Kayama_2022}.
We find that, in the region between the head and the lenticular knots, the observed data are well reproduced by assuming magnetic field strengths of 16~$\mu$G, 9~$\mu$G, and 7~$\mu$G for jet velocities of $v_{\rm jet}$ = 0.26$c$, 0.10$c$, and 0.065$c$, respectively.
However, the spectral variation in the eastern lobe cannot be fully explained by the same model used for the western lobe.
%The emission at the lenticular knot is locally brighter than in the surrounding regions, and the spectral slope gradually, rather than suddenly, steepens as one moves away from the lenticular knot along the axis.
The emission at the lenticular knot is locally brighter than in the surrounding regions.
As one moves away from the lenticular knot along the jet axis, the spectral slope steepens gradually, rather than suddenly steepens.
If the particle injection rate is constant over time and no re-acceleration occurs, a locally enhanced magnetic field is required at the knot, since synchrotron emissivity depends on both the magnetic field strength and the electron spectrum.
Even if the maximum energy of the injected electrons is higher than assumed, the spectrum would still steepen immediately outside the knot, as the cutoff energy is not sensitive to this parameter.
Therefore, we conclude that the spectral variation observed in the eastern lobe cannot be easily explained by the same model applied to the western lobe.

\subsection{Difference between the Eastern and Western Lobe}
One possible explanation for the differences in spectral variations between the western and eastern lobes is that the particles are not only cooled in the lenticular and w2 knots but also undergo other processes.
Here, we consider particle re-acceleration as an alternative scenario that could simultaneously explain both the locally bright structures and the flatter spectrum, as suggested by \cite{Kayama_2022}.
By assuming different re-acceleration efficiencies in the w2 and lenticular knots, the observed difference in spectral variation between the eastern and western lobes can be qualitatively explained.
Recently, the H.E.S.S. Collaboration reported the detection of VHE $\gamma$-rays emission from both lobes (\citealt{HESS_2024}).
The $\gamma$-ray emissions in the energy band of 2.5--10~TeV were detected around the lenticular knot, while emissions above 10~TeV were detected only in the head knot of the eastern lobe.
These results suggest that the energy of the re-accelerated particles in the lenticular knot is not as high as in the head region.
Based on the morphology of the X-ray image, it seems unlikely that diffusive shock acceleration (DSA) is the acceleration mechanism in this case, due to the absence of shock-like structures.
Instead, particle acceleration may be occurring in the knot via other mechanisms, such as Second-Order Fermi Acceleration.
Future X-ray observations in the high-energy bands will allow for a more quantitative comparison of the maximum energy differences between these knots and may provide constraints on the acceleration and cooling mechanisms.

Another possible explanation is that the same model cannot explain the situation on the eastern side as it does on the western side, because the time-evolution model of synchrotron emission used in this study is too simple to quantitatively describe the SS~433 jet.
Our model is a simple one-dimensional model that assumes a conical jet with constant velocity, but it successfully reproduces the global trend of spectral variation along the jet.
It is conceivable that the jet velocity changes as it propagates through the medium, and that the jet cross-section $S(z)$ changes due to interactions with the environment, where $z$ is the distance from the base of the jet.
In the jet, the relationship between density ($\rho$), jet velocity ($v_{\rm s}$), and $S(z)$ is given by $\rho \times v_{\rm s} \times S(z) = {\rm const}$, meaning that the compression factor $\phi$ follows $\phi \propto (v_{\rm s} \times S(z))^{-2}$ (see. equation A2 of \citealt{Kayama_2022}).
Assuming that the jet is conical with constant velocity $\phi \propto z^{-2}$, which means that the emission becomes fainter as one moves away from the base of the jet.
If the velocity is locally decelerated or if the jet is not conical, $\phi$ would increase.
The magnetic field will be locally enhanced due to plasma compression, and the adiabatic cooling/heating rate will also vary (see equation A5 of \citealt{Kayama_2022}).

Based on this discussion, the parameters of the jet on the eastern side may undergo more complex variations compared to those on the western side, which could lead to differences in spectral variation between the two jets.
In this case, assuming that the magnetic field is enhanced by the local compression of the jet plasma, the brightness can increase while maintaining a harder spectrum.
Model simulations that account for morphological changes in the jet and particle re-acceleration at the knot are beyond the scope of this study.
Future detailed simulations will provide a more robust and quantitative understanding of particle acceleration in the SS~433 jet.

\section{Conclusions}
To understand the origin of the non-thermal emission from the W~50 lobes and the relationships between them and the SS~433 jet, we performed both imaging and spectral analysis.
We found that the synchrotron X-ray spectra steepen as one moves away from the head knots, which can be explained by the synchrotron cooling of the electrons during their transport along the jet.
However, the spectrum at the lenticular knot, which is apparently equivalent to knot w2 with respect to SS~433 in the eastern lobe, maintains an almost unchanged slope.
This suggests that it cannot be explained by the same mechanism as on the western side.
One possibility is that particle re-acceleration occurs at the lenticular and w2 knots with different acceleration efficiencies between the knots.
If particles in the lenticular knot are re-accelerated with an efficiency higher than that in knot w2, this could explain both the compact, bright knots resulting from a stronger magnetic field and the flat spectrum.
Another possibility is that the simple one-dimensional model employed in our study does not fully account for the complex variation in jet shape.
Local variations in jet velocity and shape, as well as plasma compression, may affect the emissivity of the synchrotron emission.

\begin{ack}
We thank the anonymous referee for helpful comments to improve this manuscript.
We would like to thank Samar Safi-Harb and Kaya Mori for helpful comments and discussion for the data analysis.
We are also grateful to Mami Machida and Haruka Sakemi for useful discussions for the formation of W~50 and molecular cloud interaction.
We also thank Takahiro Sudoh for fruitful discussions in the calculations of the emission evolution model in the jet.
This research has made use of data obtained with the Chandra and XMM-Newton satellites.
This work is supported by Grant-in-Aid for Japan Society for the Promotion of Science (JSPS) Fellows Number JP20J23327 (K.K.). 
This work is supported also by JSPS Scientific Research Grant Nos. JP19H01936 (T.T.), JP21H04493 (T.T. and T.G.T.), JP19K03915 (H.U.), JP18H05458 (Y.I.), JP19K14772 (Y.I.), and JP22K14064 (N.T.). 
Y.I. is also supported by World Premier International Research Center Initiative (WPI), Ministry of Education, Culture, Sports, Science and Technology, Japan. 
D.K. is supported by RSF grant No.24-12-00457.
\end{ack}

\end{document}